\documentclass[prb,twocolumn,showpacs,preprintnumbers,superscriptaddress]{revtex4-1}
\usepackage[usenames,dvipsnames]{color}
\usepackage{amssymb} 
\usepackage{times}
\usepackage{graphicx}
\usepackage{graphicx}
\usepackage{dcolumn}
\usepackage{datetime}
\usepackage{soul}
\usepackage{subfigure}
\usepackage[bookmarks, colorlinks=true, plainpages = false,citecolor = red, urlcolor = black, filecolor = blue]{hyperref}
\usepackage{multirow} 
\usepackage[english]{babel}
\usepackage{graphicx}
\usepackage{amsmath}
 
\begin{document}
\title{Ligand-based transport resonances of single-molecule magnet spin
filters: Suppression of the Coulomb blockade and determination of the
orientation of the magnetic easy axis}
 
\author{Fatemeh Rostamzadeh Renani} 
\affiliation{Department of Physics, Simon Fraser University, Burnaby, British Columbia, Canada V5A 1S6}

\author{George Kirczenow} 
\affiliation{Department of Physics, Simon Fraser
University, Burnaby, British Columbia, Canada V5A 1S6}

\date{\today}

\begin{abstract}\noindent
We investigate single molecule magnet transistors  (SMMTs) with ligands that support transport 
resonances. We find the lowest unoccupied molecular orbitals of Mn$_{12}$-benzoate SMMs (with 
and without thiol or methyl-sulfide termination) to be on ligands, the highest occupied molecular 
orbitals being on the Mn$_{12}$ magnetic core. We predict  gate controlled switching between 
Coulomb blockade and coherent resonant tunneling in SMMTs based on such SMMs, strong 
spin filtering by the SMM in both transport regimes, and that if such switching is observed then the 
magnetic easy axis of the SMM is parallel to the direction of the current through the SMM.

\end{abstract}
\maketitle

Charge and spin transport through individual molecules bridging a pair of electrodes, and through 
single--molecule transistors that include a third ``gate" electrode, have been studied intensively for 
more than a decade.\cite{review2010} Because of their large magnetic anisotropy barriers and 
associated stable magnetic moments,\cite{SMMbookGatteschi} single 
molecule magnets (SMMs) raise the possibility of molecular spintronic devices
and molecular magnetic information storage.\cite{SMM2008} 
Therefore, at present, the transport properties of transistors based on individual SMMs are 
attracting considerable experimental \cite{Heersche2006,moonHo2006, Zyazin2010} and 
theoretical\cite{{Heersche2006, moonHo2006, Zyazin2010, SpinHamTheories, 
SpinHamTimmMagField2007, spinFilterBarraza09, Michalak2010, Barraza2010, GgaLdaSanvito}} 
interest.\cite{SMM2008}  
 Models based on effective spin--Hamiltonians\cite{{Heersche2006, moonHo2006, Zyazin2010, 
SpinHamTimmMagField2007,  SpinHamTheories}} and density functional theory (DFT)-based 
calculations\cite{spinFilterBarraza09, Michalak2010, Barraza2010, GgaLdaSanvito} have yielded 
many important insights into the transport properties of these systems.
However, 
in the theoretical transport studies to date the organic ligands that surround the magnetic cores of 
the SMMs have received limited attention: The focus has been on SMMs in which the lowest 
unoccupied molecular orbital (LUMO) and the highest occupied molecular orbital (HOMO) are both 
on the SMM's magnetic core. Thus the ligands have acted as simple tunnel barriers that hinder 
electron transmission between the electrodes and the magnetic core. In the case of weak 
tunneling, the electrostatic energy associated with charging of the core 
states during transport suppresses 
electrical conduction at low source-drain and gate voltages.\cite{review2010,SMM2008} This 
phenomenon is known as Coulomb blockade and has been observed in SMM transport 
experiments.\cite{Heersche2006,moonHo2006, Zyazin2010}

In this paper we explore theoretically the electron and spin transport in SMMs in which we predict the 
ligands to play a more active and interesting role: We consider Mn$_{12}$-benzoate 
[Mn$_{12}$O$_{12}$(O$_{2}$CC$_{6}$H$_{5}$)$_{16}$(H$_{2}$O$)_{4}$,  abbreviated 
Mn$_{12}$-Ph] SMMs with and without terminating methyl--sulfide or thiol groups. For these SMMs 
we predict the lowest unoccupied molecular orbital (LUMO), and orbitals nearby in energy, to be 
{\em on ligands}, and the highest occupied molecular orbital (HOMO) and orbitals nearby in energy 
to be on the SMM's magnetic core. We also predict that, for certain geometries of these SMMs 
thiol-bonded to gold electrodes, some molecular orbitals that are close in energy to the LUMO
hybridize strongly with the 
gold electrodes. It follows that for these bonding geometries transport via these near-LUMO  
orbitals should {\em not} be subject to Coulomb blockade, unlike transport via the HOMO. As we 
explain below SMM transistors in which such bonding geometries are realized can readily be 
identified experimentally by carrying out measurements of the electric current versus source-drain and 
gate voltage such as are routinely performed in single-molecule transistor experiments today. 
Furthermore, for SMM transistors identified in this way we predict the molecular easy axis to be 
approximately parallel to the direction of current flow. Thus the present theory provides a {\em 
practical} way to use standard {\em non-spin-resolved} current measurements to determine the 
orientation of the SMM magnetic easy axis, over which there was no control in SMM transistor 
experiments to date. Moreover, we predict these {\em magnetically oriented} SMM transistors  to 
be {\em effective spin filters} at low and moderate source-drain voltages. 

The results that we present are based on the semi-empirical extended H\"{u}ckel tight-binding 
model of quantum chemistry that we generalize in this paper to 
include spin polarization and spin-orbit coupling. For Mn$_{12}$ SMMs (both 
neutral molecules and negatively charged ions) the present 
model yields results consistent with experiment for the total spin of the SMM, for the spins of the 
individual Mn ions, for the direction of the magnetic easy axis, for the size of the magnetic 
anisotropy barrier (MAB), for the size of the molecular HOMO-LUMO gap and for the spins of the 
HOMO and LUMO states.\cite{othermolecs} The overall degree of agreement with experiment 
obtained with the present model is comparable to or better than that achieved with DFT 
calculations corrected by inclusion of the adjustable Hubbard $U$ 
parameter.\cite{spinFilterBarraza09, GgaLdaSanvito} However, calculations based on the present 
model are much less compute intensive than those based on DFT. Thus, we are able to study 
transport in larger molecules than are readily accessible to DFT computations and therefore,
unlike in previous theoretical studies, to include complete sets of ligands none of which have been 
shortened or replaced by hydrogen atoms.

Our SMM Hamiltonian is $H^{\mbox{\scriptsize{SMM}}}=H^{\mbox{\scriptsize{EH}}}+H^{\mbox
{\scriptsize{spin}}}+H^{\mbox{\scriptsize{SO}}}$.
Here $H^{\mbox{\scriptsize{EH}}}$ is the extended H\"{u}ckel Hamiltonian.\cite
{review2010,huckel_off_diagonal,YAEHMOP} The spin Hamiltonian $H^{\mbox{\scriptsize{spin}}}
$ gives rise to the magnetic polarization of the molecule. Spin orbit coupling is described by $H^
{\mbox{\scriptsize{SO}}}$.

In extended H\"{u}ckel theory the basis is a small set of Slater-type atomic valence orbitals $|
\Psi_{i\alpha} \rangle$;  $|\Psi_{i\alpha} \rangle$ is the $i^\mathrm{th}$ orbital of the $\alpha^
\mathrm{th}$ atom. The diagonal elements of $H^{\mbox{\scriptsize{EH}}}$ are the 
experimentally determined negative valence orbital ionization energies $\varepsilon_{i}$, $
\langle\Psi_{i\alpha} | H^{\mbox{\scriptsize{EH}}}|\Psi_{i\alpha} \rangle =H^{\mbox{\scriptsize
{EH}}}_{i\alpha;i\alpha}=\varepsilon_{i \alpha}$. The non-diagonal matrix elements are assumed 
to be proportional to the orbital overlaps $D_{i\alpha;{i}' {\alpha}'} = \langle \Psi_{i\alpha} |\Psi_
{{i}' {\alpha}'} \rangle$, i.e., $H^{\mbox{\scriptsize{EH}}}_{i\alpha;{i}' {\alpha}'}= D_{i\alpha;
{i}' {\alpha}'} K \frac{\varepsilon_{i\alpha}+\varepsilon_{{i}' {\alpha}'}}{2}$, where $K$ is chosen 
empirically for
consistency with experimental molecular electronic structure data. In our calculations,\cite
{YAEHMOP} as in Ref. \onlinecite{huckel_off_diagonal},    $K=1.75+\Delta_{i\alpha;{i}' {\alpha}'}^
{2}-0.75\Delta_{i\alpha;{i}' {\alpha}'}^{4}$ where  $\Delta_{i\alpha;{i}' {\alpha}'}=\frac{\varepsilon_{i
\alpha}-\varepsilon_{{i}' {\alpha}'}}{\varepsilon_{i\alpha}+\varepsilon_{{i}' {\alpha}'}}$. 
 
 For non-magnetic systems, transport calculations based on extended H\"{u}ckel  theory 
have yielded elastic\cite{Authioltheory, Cardamone08} and inelastic\cite{Demir2011} 
conductances in agreement with experiment for molecules
thiol-bonded to gold electrodes and have also explained 
transport phenomena observed in scanning tunneling microscopy
 (STM) experiments on molecular arrays
on silicon\cite{PivaWolkowKirczenow} as well as
electroluminescence data\cite{Buker08}, current-voltage
characteristics\cite{Buker08} and STM images\cite{Buker05} of molecules on
complex substrates.

The extended H\"{u}ckel Hamiltonian $H^{\mbox{\scriptsize{EH}}}$ does not describe 
spin polarization whereas in the Mn$_{12}$ SMMs the four inner and eight outer Mn ions
are  spin polarized with antiparallel spins. 
In our model, $H^{\mbox{\scriptsize{spin}}}$ addresses this issue. 
We define its matrix elements $\langle i s \alpha |H^{\mbox{\scriptsize{spin}}}| {i}' s' {\alpha}' 
\rangle = H^{\mathrm{spin}}_{i s \alpha;{i}'{s}' {\alpha}'}$ between
valence orbitals $i$ and $i'$ of atoms ${\alpha}$ and ${\alpha}'$ with spin $s$ and $s'$ by 
{\scriptsize\begin{equation}\label{SpinHamiltonian} \begin{split}
 &H^{\mathrm{spin}}_{i s \alpha;{i}' {s}' {\alpha}'}={D_{i \alpha;{i}'{\alpha}'}}(\mathcal{A}_{i \alpha}+
\mathcal{A}_{i'{\alpha}'}) \langle s | \hat{n}\cdot {\bf S} | {s}' \rangle/(2\hbar) \\
 &\mathcal{A}_{i \alpha}={\small \left\{\begin{matrix}
 \mathcal{A}_{\mbox{\scriptsize{inner}}} &\mbox {for inner Mn {\em d}-valence orbitals} \\ 
 \mathcal{A}_{\mbox{\scriptsize{outer}}} &\mbox {for outer Mn {\em d}-valence orbitals} \\ 
 0 &\mbox {otherwise} 
\end{matrix}\right.}
\end{split} \end {equation}}
Here $\hat{n}$ is a unit vector aligned with the magnetic moment of the SMM, and  $\bf{S}$ is 
the one-electron spin operator. $\mathcal{A}_{\mbox {\scriptsize{inner}}}$ and $\mathcal{A}_
{\mbox {\scriptsize{outer}}}$ are parameters chosen so that 
in the Mn$_{12}$ ground state the spin of each inner (outer) Mn is $S_\mathrm{inner(outer)}\cong
-\frac{3}{2}$ (+2).

Spin-orbit coupling is also not included in extended H\"{u}ckel theory.\cite
{review2010,huckel_off_diagonal,YAEHMOP} However, it is responsible for the magnetic anisotropy of 
SMMs. We therefore generalize extended H\"{u}ckel theory to include spin-orbit coupling by 
evaluating approximately the matrix elements of the spin-orbit coupling Hamiltonian $H^{\mbox
{\scriptsize{SO}}}$ starting from the standard expression\cite{Kittel}
$H^{\mbox{\scriptsize{SO}}}=\boldsymbol{\sigma}\cdot  \nabla{V(\bf{r}) \times \mathbf{p}  }~
{\hbar}/{(2mc)^2}$
where $\mathbf{p}$ is the momentum operator, $V(\bf{r})$ is the electron Coulomb potential 
energy, $\boldsymbol{\sigma} = (\sigma_x , \sigma_y, \sigma_z) $ and $\sigma_x , \sigma_y $ 
and $\sigma_z$ are the Pauli spin matrices. 
We approximate $V(\bf{r})$ by a sum of atomic potential energies $V(\mathbf{r}) \simeq \sum_
{\alpha}{V_ \alpha(\mathbf{r}-\mathbf{r}_\alpha)}$ where $\mathbf{r}_ \alpha $ is the position of $
\alpha ^\mathrm{th}$ atomic nucleus. 
Noting that the spin-orbit coupling arises mainly from the atomic cores where the potential energy 
$V_ \alpha (\bf{r}-\bf{r}_ \alpha)$ is approximately spherically symmetric, yields
$H^{\mbox{\scriptsize{SO}}} \simeq \underset {\alpha}\sum{\frac{1}{2m^2c^2} \frac{1}{|\bf{r}-\bf{r}
_ \alpha |}\frac{dV_ \alpha (|\bf{r}-\bf{r}_ \alpha |)}{d |\bf{r}-\bf{r}_ \alpha |}}\mathbf{S}\cdot \mathbf
{L}_{\alpha}$
where $\mathbf{S}=\hbar\boldsymbol{\sigma}/2$, ${\bf L}_ \alpha = (\bf{r}-\bf{r}_ \alpha) \times 
{\bf p}$ is the orbital angular momentum operator with respect to the position of the $\alpha^
\mathrm{th}$ nucleus and the sum is over all atoms $\alpha $. Evaluating the matrix elements of 
$H^{\mbox{\scriptsize{SO}}}$ between
valence orbitals $i$ and $i'$ of atoms ${\alpha}$ and ${\alpha}'$ with spin $s$ and $s'$ we then 
find
{\scriptsize
\begin{equation} 
\begin{split}
\label{SOHamiltonianfinal}
\langle {i s \alpha} & | H^{\mbox{\scriptsize{SO}}}  | {{i}' s' {\alpha}'} \rangle \simeq 
 E^{\mbox{\scriptsize{SO}}}_{is{i}'s';\alpha}\delta_{\alpha \alpha'} \\
& +(1-\delta_{\alpha \alpha'})\sum_{j}(D_{i \alpha; j\alpha'}E^{\mbox{\scriptsize{SO}}}_{js i's'; 
\alpha'} 
 +[D_{i' \alpha'; j\alpha}E^{\mbox{\scriptsize{SO}}}_{js' is; \alpha}]^{\ast}).
\end{split}
\end{equation}}The first term on the right-hand side is the intra-atomic contribution, the remaining terms are 
the inter-atomic contribution and
{\scriptsize
{\begin{equation}   
\begin{split}
 \label{factor}
E^{\mbox{\scriptsize{SO}}}_{is{i}'s';\alpha}& =  \langle \alpha, l_i, d_i, s | \mathbf S \cdot \mathbf 
L_{\alpha} | \alpha, l_{i}, d_{i'}, s' \rangle  \\
& \times \langle R_{\alpha, l_i}  | \frac{1}{2m^2c^2} \frac{1}{|\mathbf{r}-\mathbf{r}_{\alpha}|} \frac
{dV(|\mathbf{r}-\mathbf{r}_{\alpha}|)}{d(|\mathbf{r}-\mathbf{r}_{\alpha}|)}  | R_{\alpha, l_{i}}  \rangle 
\delta_{l_i l_{i'}}
\end{split}
 \end{equation}}}where the atomic orbital wave function $\Psi_{i \alpha}$   has been expressed 
as the product of a radial wave function $R_{\alpha, l_i}$ and directed atomic orbital $|\alpha, l_i, 
d_i, s \rangle$. Here $l_i$ is the angular momentum quantum number and $d_i$ may be $s, p_x, 
p_y, p_z, d_{xy},d_{xz},...$ depending on the value of $l_i$. In Eq. (\ref{factor}) the matrix 
elements of ${\bf S} \cdot \mathbf L_{\alpha}$  are evaluated as in Ref. \onlinecite
{SpinOrbiGraphene} while 
the radial integrals are the spin-orbit coupling constants.

We estimate the magnetic anisotropy energy of the SMM from the total ground state energy 
expression
$E_{\mbox{\scriptsize{total}}}=\sum_{i}^{}E_{i} - \sum_{i}^{}\left \langle \phi_{i} \left | H^{\mbox
{\scriptsize{spin}}} \right | \phi_{i} \right \rangle/2$
where $E_{i}$ and $|\phi_{i}\rangle$ are eigenenergies and eigenstates of $H^{\mbox{\scriptsize
{SMM}}}$ and the summations are over all occupied states. Since $H^{\mbox{\scriptsize{spin}}}$ 
represents electron-electron interactions at a mean field level, the second summation on the right 
hand side is required to avoid double counting the corresponding interaction energy.  

Experimental estimates\cite{SOMnExperiment} of the spin-orbit coupling constant for Mn $d$-orbitals 
have been in the range 0.023--0.051$~$eV, while theoretical estimates\cite{SOMnTheory} 
have been in the range 0.038--0.055$~$eV. In this paper for Mn atoms we use the value 
0.036$~$eV, which is consistent with the experimental and theoretical values. We find that the 
spin-orbit coupling constants of the other atoms do not affect the SMMs' properties significantly 
and that the intra- and inter-atomic terms in Eq. (\ref {SOHamiltonianfinal}) make contributions of 
the same order of magnitude to the magnetic anisotropy barriers of Mn$_{12}$ SMMs. In this 
work the values of  $H^{\mbox{\scriptsize{EH}}}_{i\alpha,i'\alpha'}$ and $D_{i\alpha,i'\alpha'}$ that 
enter the extended H\"{u}ckel model, Eqs. (\ref{SpinHamiltonian}) and (\ref 
{SOHamiltonianfinal}), were adopted without change from  Refs. \onlinecite{huckel_off_diagonal} 
and \onlinecite{YAEHMOP}.  The molecular geometries that we studied were based on the 
experimentally measured geometry\cite{Geometry_data_Et_and_Ph} of Mn$_{12}$-Ph modified 
as necessary by adding thiol or methyl sulfide groups to the ligands. Thus the only free 
parameters in the present theory are the $\mathcal{A}_{i \alpha}$ of Eq. (\ref{SpinHamiltonian}) 
that control the spin polarizations of the Mn atoms; we chose $\mathcal{A}_{\mbox{\scriptsize
{inner}}}=3.0~$ eV and $\mathcal{A}_{\mbox{\scriptsize{outer}}}=-3.5~$ eV. 
For these parameter values in the Mn$_{12}$-Ph ground state we find the inner 
and outer Mn ions to have spins --1.6 and 1.99, the SMM to have a total spin of 10 
and the calculated 
MAB to be 2.50 meV, all consistent with experiment. \cite{Geometry_data_Et_and_Ph, 
MABneutralMn12Ph} 

\begin{figure}[t!] 
\centering
\includegraphics[width=0.90\linewidth]{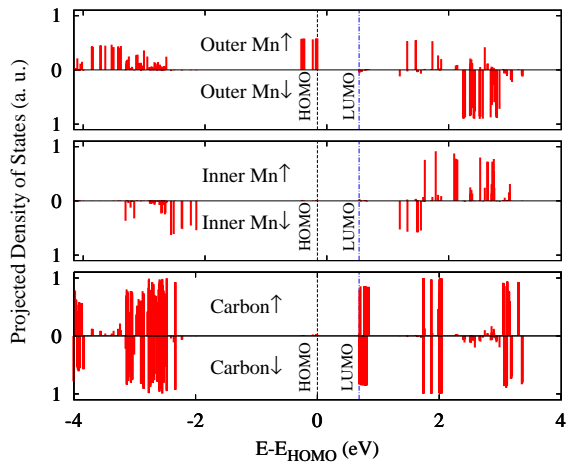} 
\caption{Projected density of states for majority (spin-up) and  minority (spin-down) 
electrons on Mn and carbon atoms
for the isolated Mn$_{12}$-Ph molecule without thiol or methylsulfide end groups.} 
\label{DosMnPh} 
\end{figure}

The calculated densities of states projected on the inner and outer Mn atoms, and carbon 
atoms 
of Mn$_{12}$-Ph are shown  in Fig. \ref{DosMnPh}.
The HOMO and nearby levels are on the outer Mn and are filled with spin-up 
electrons (parallel to the total spin), consistent with $S_{\mbox{\scriptsize
{outer}}}=+2$. The occupied inner Mn states are filled with spin-down electrons  consistent with 
$S_{\mbox{\scriptsize{inner}}}=-\frac{3}{2}$. The carbon atoms are weakly spin polarized. 
The calculated HOMO-LUMO energy gap is $\sim$0.7eV. 
(Note that experimental and theoretical estimates for the Mn$_
{12}$ family range from a few tenths of an electron Volt
to more than 1 eV.\cite{GgaLdaSanvito, Barraza2010, MnAcExperimental_HomoLumoGap})
As is seen in Fig. \ref{DosMnPh}, our calculation predicts the LUMO of Mn$_{12}$-Ph to be 
mainly on the carbon atoms of the ligands. There is at this time
no {\em direct} experimental evidence as to whether the LUMO of Mn$_{12}$-Ph is located on 
the ligands as in Fig. \ref{DosMnPh} or on the Mn$_{12}$ core of the molecule. However, 
experimental measurements of the MAB carried out on the neutral\cite{MABneutralMn12Ph} 
and 
negatively charged\cite {MABchargedMn12Ph} Mn$_{12}$-Ph species yielded values of  3.3  
and 2.41 meV, respectively, a difference of only $27\%$. Since the main source of the MAB of 
Mn
$_{12}$ is the Jahn-Teller distortion of the Mn$_{12}$ core of the molecule\cite{moonHo2006} 
this insensitivity of the MAB to the oxidation state of the molecule is consistent with the added 
electron of the negatively charged species residing mainly on the ligands rather 
than on the Mn$_
{12}$ core, as one might expect if the LUMO of the neutral molecule is on the 
ligands 
as in Fig. \ref{DosMnPh}. Our calculations also predict only a small change in the MAB when 
an 
electron is added to the neutral Mn$_{12}$-Ph molecule, since in our model the added electron 
locates primarily on the ligands rather than the magnetic molecular core. Our prediction that the
 LUMO is on the ligands(i.e. the benzoate groups, including both their carbon and oxygen atoms) 
 and not on the Mn$_{12}$ core is also consistent with the large 3.3--3.6 eV 
electron affinity of the benzoate species. \cite{BenzoateEA, footnote}

\begin{figure}[b!]
\centering
\includegraphics[width=1\linewidth]{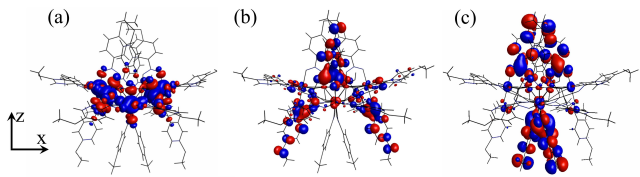}
\caption{Wave functions for Mn$_{12}$-Ph-Th (a) spin-up HOMO (b) spin-down LUMO (c) spin-up 
orbital near LUMO in energy. The magnetic easy axis and total molecular spin are parallel to the $z$--
axis. Although the LUMO and near LUMO orbitals are mainly on the ligands they are spin polarized 
due to their small but non-zero overlaps with the magnetic core of the molecule that are visible in (b) and (c).}
\label{MolecularOrbital}
\end{figure}

\begin{figure}[b]
\centering
\includegraphics[width=1.0\linewidth]{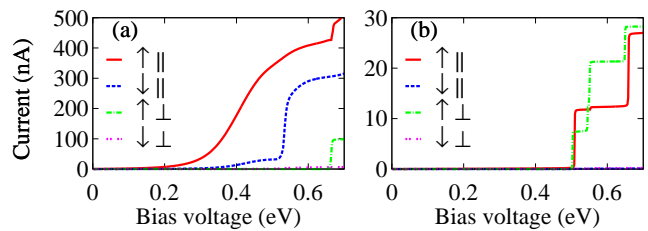}
\caption{Calculated spin resolved current  parallel ($\parallel$) and perpendicular ($\perp$) to the
easy axis for (a) positive and (b) negative gate voltage vs. bias 
voltage at zero temperature. The gate potential {\em at the molecule} = +0.2 and --0.2V in (a) and 
(b), respectively. }
\label{TransmissionMnPhSOP}
\end{figure}

The ligands of SMMs that are studied in transport experiments with gold electrodes are normally 
thiolated and we shall therefore focus our attention on Mn$_{12}$-Ph-Th, i.e., Mn$_{12}$-Ph 
terminated with methyl sulfide (SCH$_3$) groups, the methyl being displaced by gold when the 
molecule bonds to the contacts. The calculated densities of states of Mn$_{12}$-Ph-Th (and
of Mn$_{12}$-Ph with thiol end groups) projected on the 
Mn and C atoms are similar to those shown in Fig.\ref{DosMnPh} for Mn$_{12}$-Ph. The locations 
and spins of the HOMO and LUMO and of orbitals close to these in energy are also similar for Mn$_{12}$-Ph
with and without thiol or methyl sulfide end groups. The calculated 
HOMO, LUMO and another representative molecular orbital close in energy to the LUMO are 
shown in Fig.\ref{MolecularOrbital} for Mn$_{12}$-Ph-Th. The HOMO is located on the magnetic core of the molecule. 
However, the LUMO and molecular orbitals close in energy to the LUMO are located on ligands, 
and specifically {\em on those ligands that are oriented approximately parallel to the magnetic easy 
axis of the molecule} which points in the $z$--direction in Fig. \ref{MolecularOrbital}. This finding has 
important implications for electron and spin transport in Mn$_{12}$-Ph-Th-based SMM transistors: 
The HOMO and molecular orbitals close in energy to the HOMO have very little overlap with 
any of the ligands. Therefore {\em all} of the ligands that couple the molecule to the contacts act as strong 
tunnel barriers for transport mediated by the HOMO and molecular orbitals 
nearby in energy. For this reason transport via the HOMO and molecular orbitals close in energy to 
the HOMO is predicted to display the classic signature of Coulomb blockade. By contrast if any gold 
contact bonds to a ligand on which certain molecular orbitals close in energy to the LUMO
(for example, that in Fig. \ref {MolecularOrbital}(c)) 
have a strong presence, those molecular orbitals will hybridize strongly with the gold contact and 
therefore transport via that molecular orbital (or orbitals) will not be subject to Coulomb blockade. 
Furthermore, because these orbitals occupy ligands that are oriented approximately parallel to the 
molecular easy axis as in  Fig. \ref{MolecularOrbital}, if such transport that is not subject to 
Coulomb blockade is observed experimentally, then the molecule must be oriented relative to the 
gold contacts in such a way that the magnetic easy axis is approximately parallel to the direction of 
current flow through the molecule. We note that although there have been theoretical suggestions 
previously of possible ways to determine the orientation of the easy axis in a SMM transistor,\cite
{SpinHamTimmMagField2007,Barraza2010} these suggestions have been difficult to implement in 
practice and no experimental control over the orientation of the easy axis in SMM transitors has 
been achieved experimentally to date. The present theory is much more promising in this regard 
since the experimental observation of the presence or absence of Coulomb blockade in single--molecule 
transistors is currently carried out routinely, and we predict that if Coulomb blockade is 
observed at negative gate voltages (transport via the HOMO and nearby states) but not at 
positive gate voltages (transport via the LUMO or nearby states) then the easy axis is 
approximately parallel to the direction of current flow. 

Our transport calculations based on Landauer theory and the Lippmann-Schwinger equation\cite
{transportmethod} predict Mn$_{12}$-Ph-Th-based SMM transistors with gold contacts to be 
effective spin filters at low source-drain bias for both positive and negative gate voltages. 
Representative results for a Mn$_{12}$-Ph-Th molecule bonded to gold contacts {\em via ligands 
on which near-LUMO orbitals have a strong presence} (and therefore the current through the SMM 
is roughly parallel to the magnetic easy axis) are shown in Fig.\ref{TransmissionMnPhSOP}. Here 
mainly spin-up electrons are transmitted through the SMM at low bias for both positive (Fig.\ref
{TransmissionMnPhSOP}(a)) and negative (Fig.\ref{TransmissionMnPhSOP}(b)) gate voltages. The 
{\em gradual} rise of the current with bias voltage (from $\sim$0.2 to $\sim$0.67V) in Fig.\ref
{TransmissionMnPhSOP}(a) is a direct manifestation of the large broadening of the near-LUMO 
molecular orbitals responsible for transport that is due to the strong hybridization of those orbitals 
with the gold contacts that also suppresses Coulomb blockade for positive gate bias. By contrast, 
the {\em abrupt} step-like rise to much lower values of the current 
for negative gate voltages (Fig.\ref
{TransmissionMnPhSOP}(b)) is due to the near-HOMO molecular orbitals being very weakly 
coupled to the gold contacts and therefore being only very weakly broadened, they are subject 
to Coulomb blockade.\cite{limitation} For molecules bonded to gold 
electrodes via ligands that are roughly 
perpendicular to the easy axis, neither near-LUMO nor near-HOMO orbitals have significant 
presence on the ligands, see Fig. \ref{MolecularOrbital}. Thus, for geometries with a current 
through the SMM that is roughly perpendicular to the magnetic easy axis, the ligands act as strong 
tunnel barriers for both positive and negative gate voltages (Fig. \ref {TransmissionMnPhSOP}). 
Coulomb blockade is therefore predicted for {\em both} signs of the gate voltage.

In conclusion, previous studies of transport in single molecule magnets have considered the 
situation where the HOMO and LUMO both reside on the magnetic core of the molecule. Here we 
have proposed that this need not always be the case, Mn$_{12}$ benzoate with and without 
terminating methyl--sulfide or thiol groups being a possible example. We have predicted that for 
these systems the LUMO and molecular orbitals close in energy to the LUMO reside on the organic 
ligands that are oriented approximately parallel to the magnetic easy axis of the molecule and that 
when the molecule bonds via these ligands to gold electrodes in a single--molecule transistor, 
transport via some of the near-LUMO orbitals should {\em not} be in the Coulomb blockade 
regime. 
For other orientations 
of the molecule transport via the LUMO and near-LUMO orbitals 
is predicted be in the Coulomb blockade regime, as is 
transport via the HOMO and near-HOMO orbitals  for all molecular orientations. 
This effect should make it possible to study experimentally the transport in single molecule magnet
transistors that behave as spin filters and in which the orientation of the 
magnetic easy axis relative to the electrodes is 
known. In  single--molecule magnet transistor experiments to date\cite{} the orientation of the magnetic easy axis 
has not been determined although it controls the spin polarization of the current in such devices. 

While this work has focused on Mn$_{12}$ benzoate and its derivatives, we expect other 
single molecule magnets with LUMO and/or HOMO states located on the ligands to exist as well 
due to the small HOMO-LUMO gaps exhibited by a variety of organic molecules that
may be chosen as ligands. For example it is well established that polyacetylene and polythiophene 
have HOMO-LUMO gaps of 1.4 eV\cite{polyacetylene} and 0.85 eV\cite{polythiophene}, respectively.
Both of these are smaller than the experimentally measured  energy gap ($\sim$ 1.8 eV)\cite{Mn12Gap} between the highest occupied and lowest unoccupied orbitals localized on the cores of the Mn$_{12}$ molecules.
It is reasonable to expect oligomers (consisting of several monomers) derived from these polymers to have
similarly small HOMO-LUMO gaps. If such oligomers are used as ligands 
for Mn$_{12}$ SMMs contacted with gold electrodes, the gold Fermi level 
is expected to lie within the oligomer HOMO-LUMO gap and also within the HOMO-LUMO gap 
of the Mn$_{12}$ core. Therefore either the HOMO or the LUMO (or both) of the Mn$_{12}$
SMM with such ligands is expected to lie on the ligands.
 
This research was supported by CIFAR, NSERC, Westgrid and Compute Canada. 
We thank  B. Gates and B. L. Johnson for helpful comments and discussions.
 

\end{document}